\documentclass[10pt]{iopart}

\usepackage{upgreek}
\usepackage{graphicx}
\usepackage{siunitx}
\usepackage{floatrow}
\usepackage{relsize}
\usepackage{gensymb}
\usepackage{xcolor}
\usepackage[square,numbers,sort&compress]{natbib}
\def\unit#1{\mathord{\thinspace\rm #1}}

\begin{document}

\title{Tailoring coercive fields and the Curie temperature via proximity coupling in WSe$_2$/Fe$_3$GeTe$_2$ van der Waals heterostructures}

\author{Guodong Ma$^{1,4}$, Renjun Du$^{1,4}$, Fuzhuo Lian$^{1,2}$, Song Bao$^1$, Zijing Guo$^1$, Xiaofan Cai$^1$, Jingkuan Xiao$^1$, Yaqing Han$^1$, Di Zhang$^1$, Siqi Jiang$^1$, Jiabei Huang$^1$, Xinglong Wu$^1$, Alexander S. Mayorov$^{1,*}$, Jinsheng Wen$^{1,3}$, Lei Wang$^{1,3,*}$ and Geliang Yu$^{1,3,*}$ }

\address{$^1$ National Laboratory of Solid State Microstructures, School of Physics, Nanjing University, Nanjing 210093, China}
\address{$^2$ Chongqing 2D Materials Institute, Liangjiang New Area, Chongqing, 400714, China}
\address{$^3$ Collaborative Innovation Center of Advanced Microstructures, Nanjing University, Nanjing 210093, China}
\address{$^4$ These authors contributed equally to this work.}
\address{$^*$ Authors to whom any correspondence should be addressed.}

\eads{\mailto{mayorov@nju.edu.cn}, \mailto{leiwang@nju.edu.cn}, \mailto{ yugeliang@nju.edu.cn}}

\vspace{10pt}
\begin{indented}
\item[]\today
\end{indented}
\vspace{10pt}
\begin{indented}
	\item[]{\it Keywords}: vdW heterostructure, 2D magnetism, spin-orbit coupling, proximity effect
\end{indented}

\begin{abstract}
Hybrid structures consisting of two-dimensional (2D) magnets and semiconductors have exhibited extensive functionalities in spintronics and opto-spintronics.
In this work, we have fabricated WSe$_2$/Fe$_3$GeTe$_2$ van der Waals (vdW) heterostructures and investigated the proximity effects on 2D magnetism. 
Through reflective magnetic circular dichroism (RMCD), we have observed a temperature-dependent modulation of magnetic order in the heterostructure. For temperatures above $40\unit{K}$, WSe$_2$-covered Fe$_3$GeTe$_2$ exhibits a larger coercive field than that observed in bare Fe$_3$GeTe$_2$, accompanied by a noticeable enhancement of the Curie temperature by $21\unit{K}$. This strengthening suggests an increase in magnetic anisotropy in the interfacial Fe$_3$GeTe$_2$ layer, which can be attributed to the spin-orbit coupling (SOC) proximity effect induced by the adjacent WSe$_2$ layers. However, at much lower temperatures ($T<20\unit{K}$), a non-monotonic modification of the coercive field is observed, showing both reduction and enhancement, which depends on the thickness of the WSe$_2$ and Fe$_3$GeTe$_2$ layers. Moreover, an unconventional two-step magnetization process emerges in the heterostructure, indicating the short-range nature of SOC proximity effects.
Our findings revealing proximity effects on 2D magnetism may shed light on the design of future spintronic and memory devices based on 2D magnetic heterostructures.
\end{abstract}

%
%
\submitto{\TDM}
%
%
\ioptwocol

\section{Introduction}

The integration of magnetic materials with semiconductors has been extensively explored in spintronics through active manipulations of the spin degrees of freedom\cite{1990Science,2004Rev-of-MP}. This exploration has led the investigation of various prototypical devices, such as spin field-effect transistors\cite{Datta-Das} and magnetic tunnel junctions\cite{2023AM10,2023AM11}, to unlock fascinating functionalities within these hybrid structures. The interfaces between magnets and semiconductors play a critical role in determining these novel properties.
Recently, the utilization of two-dimensional materials to form van der Waals (vdW) magnetic heterostructures has provided a new platform for the development of ultrathin spintronic devices\cite{2020NPJreview,2021NNreview,2021APRreview,2023AMreview}, owing to their high-quality crystalline structures and bond-free interfaces\cite{Kreview2016, Duanreview2016,2dapp3}.

Incorporating 2D magnets with 2D semiconductors in a vdW heterostructure enables a wide range of intriguing phenomena through magnetic proximity effects.
For instance, WSe$_2$/CrI$_3$ heterostructures demonstrate a spontaneous emergence of both valley Zeeman splitting and polarization\cite{2017WSe-CrI,2020WSe-CrI}. The former is highly sensitive to the overall magnetization while the latter is dominated by the interfacial layer of CrI$_3$ and can be controlled through electrical gating\cite{Gating-theory,pr-material}. Additionally, the magnetic proximity effect in MoSe$_2$/CrBr$_3$ exhibits a dependency on the charge state\cite{2020MoSe-CrBr}. The examples mentioned above focus on the impact of magnetic layers on nonmagnetic semiconductors. 
Conversely, there is growing interest in modulating 2D magnets through adjacent nonmagnetic layers\cite{2021ACSnano,2022MoS2-FGT, WS2-FePS3,2023ACSnano}. In 2D magnets, spin-orbit coupling (SOC) from heavy elements is crucial for magnetic anisotropy, which sustains long-range magnetic order even at the monolayer limit\cite{2dmag_CrI3,fgt2018}. It has been proposed that combining 2D magnets with materials that exhibit strong SOC can effectively enhance the magnetic anisotropy\cite{SongC-PRB17, SongC-PRB20, W-CGT, CGT-SOC,2023ACSnano}, resulting in a modification of the magnetic order. However, experimental studies are few, which require extensive efforts.

Here, we investigate the influence of proximity effects on a magnetic layer in vdW heterostructures, consisting of a 2D itinerant magnet Fe$_3$GeTe$_2$ (FGT) and a 2D semiconductor WSe$_2$ with strong SOC\cite{WSe-SOC-theory,2021NNreview}. Through reflective magnetic circular dichroism (RMCD) studies, we observe a pronounced enhancement of the coercive field and Curie temperature in the WSe$_2$/Fe$_3$GeTe$_2$ heterostructure compared with bare FGT, arising from SOC occurring at the interface.
However, a non-monotonic tuning of coercive fields, including both reduction and enhancement, emerges in WSe$_2$/Fe$_3$GeTe$_2$ regions at much lower temperatures ($<20\unit{K}$), depending on the thickness of WSe$_2$ and Fe$_3$GeTe$_2$ layers.
Moreover, an unconventional magnetic stratification is observed in the heterostructure with monolayer WSe$_2$, leading to an obvious two-step magnetization reversal, which reveals the short-range nature of SOC proximity effects.

\section{Results and discussions}

Figure 1a shows the schematic of the WSe$_2$/Fe$_3$GeTe$_2$ heterostructure, assembled by an all-dry viscoelastic stamping method\cite{PDMS_method}.  
In our experiments, we exfoliated Fe$_3$GeTe$_2$ flakes onto a Si substrate with $285\unit{nm}$ SiO$_2$. The thin WSe$_2$ flakes were exfoliated on polydimethylsiloxane (PDMS). Then we transferred the WSe$_2$ onto the Fe$_3$GeTe$_2$ flake via a high-precision transfer setup to construct the final heterostructure. The fabrication was carried out inside a glovebox with an argon atmosphere to avoid the reaction with oxygen and moisture. Figure 1b shows an optical image of the heterostructure H1 with few-layer FGT and bilayer WSe$_2$. The thickness of the FGT flake is $6.2\unit{nm}$ (about 7 layers) determined by atomic force microscopy (AFM) (see the inset of Fig. 1b). The room-temperature photoluminescence (PL) spectra of the WSe$_2$ flakes (Fig. 1c) exhibit a distinct dependence on thickness, in which a strong single peak centered at $1.67\unit{eV}$ emerges in the monolayer WSe$_2$\cite{WSe-PL}. This allows us to identify the layer numbers of the corresponding WSe$_2$ flakes in heterostructures.

 \begin{figure*}[htb]
	\includegraphics[width=1\textwidth]{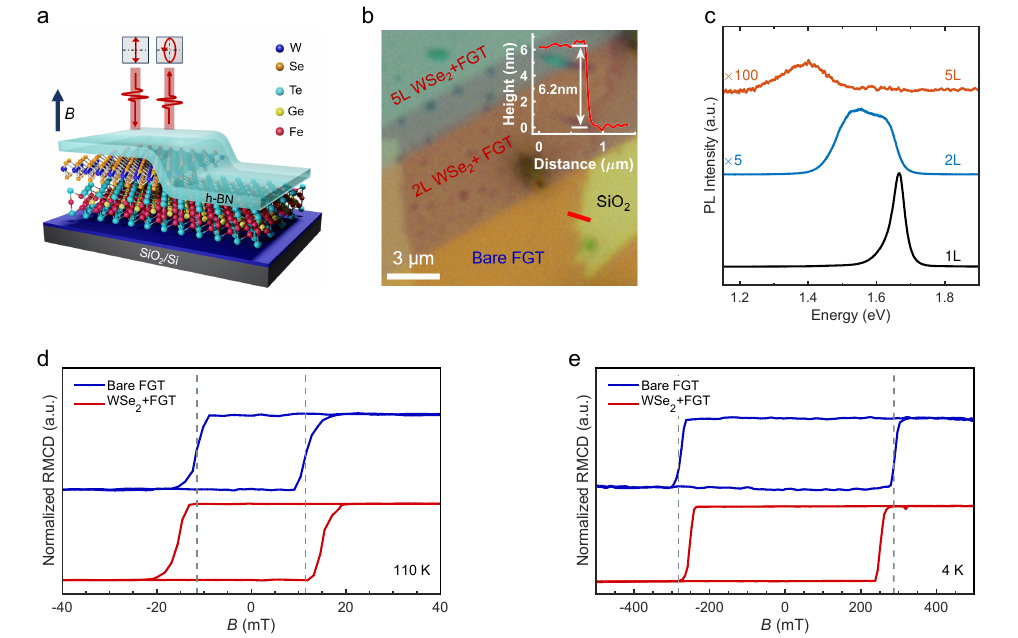}
	\caption{(a) Schematic of the RMCD measurement configuration and crystal structure of the WSe$_2$/Fe$_3$GeTe$_2$ heterostructure. (b) Optical image of bilayer WSe$_2$ on Fe$_3$GeTe$_2$. The inset shows the profile extracted along the red line in (b). (c) PL spectra of the WSe$_2$ flakes with different thicknesses. (d,e) Comparison of normalized RMCD signal for bare FGT and WSe$_2$-covered FGT in H1 at (d) $110\unit{K}$ and (e) $4\unit{K}$, respectively. The grey dashed lines mark the positions of coercive fields of bare FGT.}
	\label{fig1}
\end{figure*}

To investigate the WSe$_2$-induced proximity effect on magnetic order, we performed a series of RMCD measurements by an optical cryostat.
A $633\unit{nm}$ HeNe laser with excitation power of $10\unit{\upmu W}$ was used and focused onto the sample (Fig. 1a) with a spot size of about $1\unit{\upmu m}$. Figures 1d,e show a comparison of the RMCD signals of  WSe$_2$-covered FGT (WSe$_2$+FGT) region and the bare FGT region from the same heterostructure H1 (bilayer WSe$_2$/6.2 nm FGT), at the temperature of $110\unit{K}$ and $4\unit{K}$, respectively. The typical rectangular hysteresis loops emerge in both regions. Notably, we observe an opposite effect on the coercive field $H_\mathrm{c}$ at different temperatures. At $110\unit{K}$, the loop in the WSe$_2$+FGT region (red curve at the bottom) is broader than that measured in bare FGT (blue curve at the top), as shown in Fig. 1d, while $H_\mathrm{c}$ exhibits a reduction in the WSe$_2$+FGT region at $4\unit{K}$.

\begin{figure*}[htb]
	\includegraphics[width=1.0\textwidth]{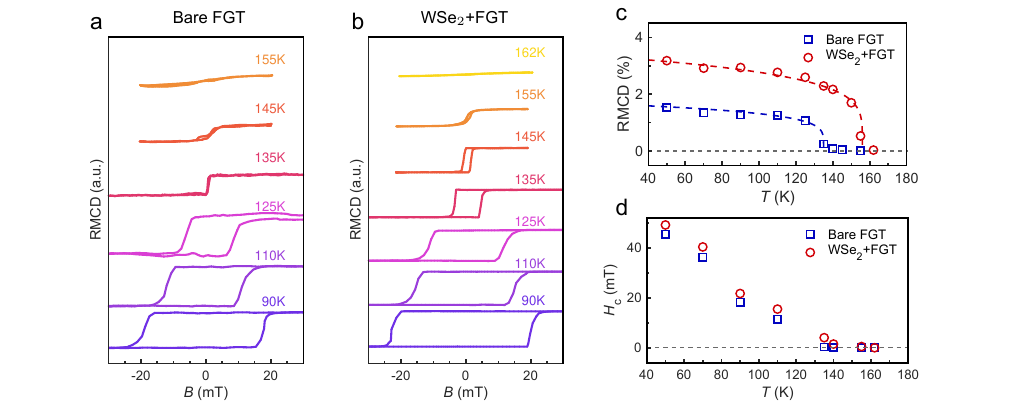}
	\caption{Proximity-induced enhancing PMA in H1. (a,b) Comparison of temperature-dependent RMCD signal for bare FGT (a) and WSe$_2$-covered FGT (b). (c) Remanent RMCD signal as a function of temperature for different regions in H1. The dashed fitting curves are plotted by the critical power-law form $\alpha(1-T/T_\mathrm{c})^\beta$. (d) Extracted coercive field $H_\mathrm{c}$ as a function of temperature for different regions in H1.}
	\label{fig2}
\end{figure*}

Figures \ref{fig2}a,b present the temperature dependence of magnetic hysteresis loops in sample H1, compared between the bare FGT and WSe$_2$+FGT regions. As temperature increases, the hysteresis loop shrinks rapidly in both regions due to the stronger thermal fluctuation. It disappears at $135\unit{K}$ in bare FGT, implying a transition from the ferromagnetic phase to paramagnetic phase, but retains a rectangular loop in the WSe$_2$+FGT region under the same conditions. Figure \ref{fig2}c shows the remanent RMCD signal as a function of temperature, in which we also compare the results for bare FGT (blue squares) and WSe$_2$+FGT (red circles). The corresponding dashed curves are plotted by the critical power-law form $\alpha(1-T/T_\mathrm{c})^\beta$, by which we extract the Curie temperature $T_\mathrm{c}$ and the critical exponent $\beta$. The $T_\mathrm{c}$ of WSe$_2$-covered FGT is increased by $21\unit{K}$ due to the WSe$_2$ proximity effect ($135\unit{K}$ in bare FGT and $156\unit{K}$ for WSe$_2$+FGT). The value of $\beta$ we obtain is $0.19\pm0.01$, consistent with the previous report on a few layers of FGT\cite{fgt2018}. Figure \ref{fig2}d displays the corresponding extracted $H_\mathrm{c}$ as a function of temperature for both two regions. The enhancement of $H_\mathrm{c}$ can be observed in the same temperature range. 
The significant increase in $T_\mathrm{c}$ and the simultaneous enhancement of $H_\mathrm{c}$ confirm the improved magnetism in FGT, which can be attributed to the proximity effect from the adjacent WSe$_2$.

\begin{figure*}[htb]
	\includegraphics[width=1\textwidth]{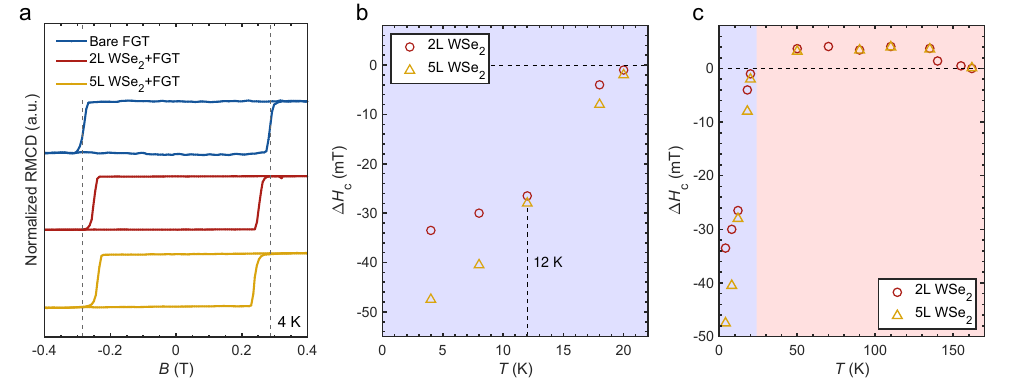}
	\caption{(a) Normalized RMCD signal measured at different regions: blue line, bare FGT; red line, 2L WSe$_2$/FGT; yellow line, 5L WSe$_2$/FGT. (b) $H_\mathrm{c}$ difference, $\Delta H_\mathrm{c}=H_\mathrm{c}^\mathrm{W}-H_\mathrm{c}^\mathrm{FGT}$, as a function of temperature, where $H_\mathrm{c}^\mathrm{W}$ and $H_\mathrm{c}^\mathrm{FGT}$ represent the coercive field in regions of WSe$_2$-covered FGT and bare FGT, respectively. (c) $\Delta H_\mathrm{c}$ as a function of temperature with the full temperature range. Two distinct background colors represent the sign of $H_\mathrm{c}$ (blue, negative; red, positive, respectively).}
	\label{fig3}
\end{figure*}

In contrast to the enhancement of coercive fields at higher temperatures ($T>40\unit{K}$), as shown in Fig. \ref{fig2}, WSe$_2$-covered FGT demonstrates a distinct reduction of $H_\mathrm{c}$ ($\sim330\unit{Oe}$) at $4\unit{K}$ (see Fig.~\ref{fig3}a). We measured the layer dependence of WSe$_2$ on $H_\mathrm{c}$, as shown in Fig. \ref{fig3}a, demonstrating a discernible reduction in $H_\mathrm{c}$ with the increase in layer numbers. 
Figures \ref{fig3}b,c depict the difference of coercive fields ($\Delta H_\mathrm{c}=H_\mathrm{c}^\mathrm{W}-H_\mathrm{c}^\mathrm{FGT}$) as a function of temperature in both 2L-WSe$_2$/FGT and 5L-WSe$_2$/FGT regions, where $H_\mathrm{c}^\mathrm{W}$ and $H_\mathrm{c}^\mathrm{FGT}$ represent the extracted coercive field in WSe$_2$-covered FGT and bare FGT, respectively. 
As shown in Fig. \ref{fig3}b, the negative $\Delta H_\mathrm{c}$ associated with the suppression of magnetic anisotropy is evident in both 2L- and 5L-WSe$_2$ covered regions  at low temperatures, persisting as the temperature increases until $20\unit{K}$. $H_\mathrm{c}$ in 5L-WSe$_2$+FGT is smaller than that in 2L-WSe$_2$+FGT, suggesting a stronger suppression of magnetic anisotropy. Additionally, as the temperature rises above $12\unit{K}$, $H_\mathrm{c}$ of the 2-layer (red circles) and the 5-layer (yellow triangles) WSe$_2$+FGT are almost equal within the margin of error. Figure \ref{fig3}c presents the results for the overall temperature range below $T_\mathrm{c}$. The sign of $\Delta H_\mathrm{c}$ changes from negative (blue panel) to positive (red panel) at approximately $20\unit{K}$, and the regions with different thicknesses of WSe$_2$ show no pronounced difference at elevated temperatures. 

Thus far, our observations have revealed the coexistence of the proximity-induced enhancement and suppression of magnetic anisotropy in FGT, suggesting the likelihood of multiple underlying mechanisms. In 2D magnets, magnetic anisotropy arising from spin-orbit coupling plays a crucial role in stabilizing magnetic order. It has been proposed that attaching a nonmagnetic layer with strong SOC to 2D magnets can effectively enhance the magnetic anisotropy\cite{CGT-SOC}.
Recent experiments involving analogous heterostructures\cite{2021ACSnano,2023ACSnano} reveal the stabilization of 2D magnetism in thick FGT flakes, which is induced by the SOC proximity effect. Therefore, it is reasonable to ascribe the enhancing magnetic anisotropy to the SOC proximity effect. On the other hand, the magnetic anisotropy of FGT can also be modulated by strain engineering\cite{FGT-PRB, FGTstrain}. Given that the lattice constant in WSe$_2$ is smaller than that in FGT, a slightly compressive strain may occur in the $ab$-plane of FGT\cite{SongC-PRB17}, potentially contributing to the suppression of magnetic anisotropy at low temperatures.

%

\begin{figure*}[htb]
	\includegraphics[width=1\textwidth]{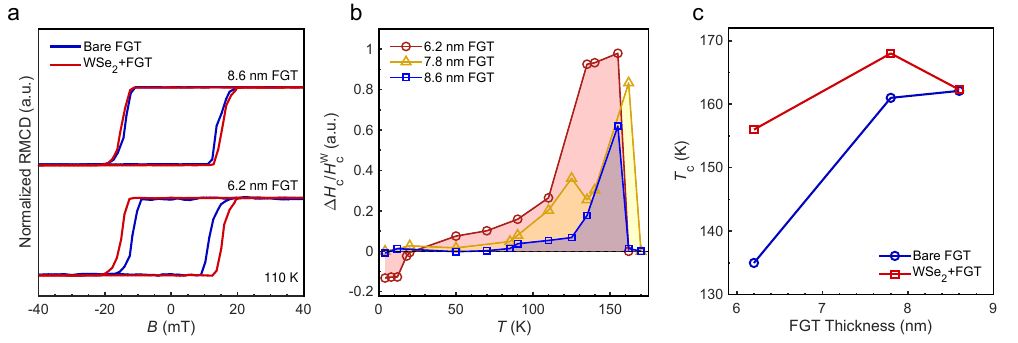}
	\caption{(a) Comparison of normalized RMCD signal in different heterostructures with FGT thicknesses of $6.2\unit{nm}$ and $8.6\unit{nm}$ at $110\unit{K}$. (b) Relative differences of the coercive field $\Delta H_\mathrm{c}/H_\mathrm{c}^\mathrm{W}$ as a function of temperature. (c) Thickness dependence of $T_\mathrm{c}$ in the bare FGT and the WSe$_2$-covered FGT.}
	\label{fig4}
\end{figure*}

Figure \ref{fig4}a shows the RMCD measurements in two WSe$_2$+FGT heterostructures with varying thicknesses of FGT but the same thickness of WSe$_2$ flakes (2 layers). The heterostructure with a thinner FGT exhibits a more pronounced enhancement of $H_\mathrm{c}$ at $110\unit{K}$. 
We present the relative difference $\Delta H_\mathrm{c}/H_\mathrm{c}^\mathrm{W}$ as a function of temperature in Fig. \ref{fig4}b. It is noteworthy that the reduction of $H_\mathrm{c}$ at low temperatures is only observable in the thinnest FGT ($6.2\unit{nm}$), but is absent in the heterostructures with thicker FGT, which implies that the potential strain effect rapidly diminishes as the FGT thickness increases.
Figure \ref{fig4}c illustrates the dependence of $T_\mathrm{c}$ on FGT thicknesses. At the thickness of $8.6\unit{nm}$, $T_\mathrm{c}$ of WSe$_2$/FGT and bare FGT are nearly identical; however, a significant difference is observed at the thickness of 6.2 nm, showing that the enhancement induced by SOC proximity is strengthened with decreasing thickness of FGT.

\begin{figure*}[htb]
	\includegraphics[width=1\textwidth]{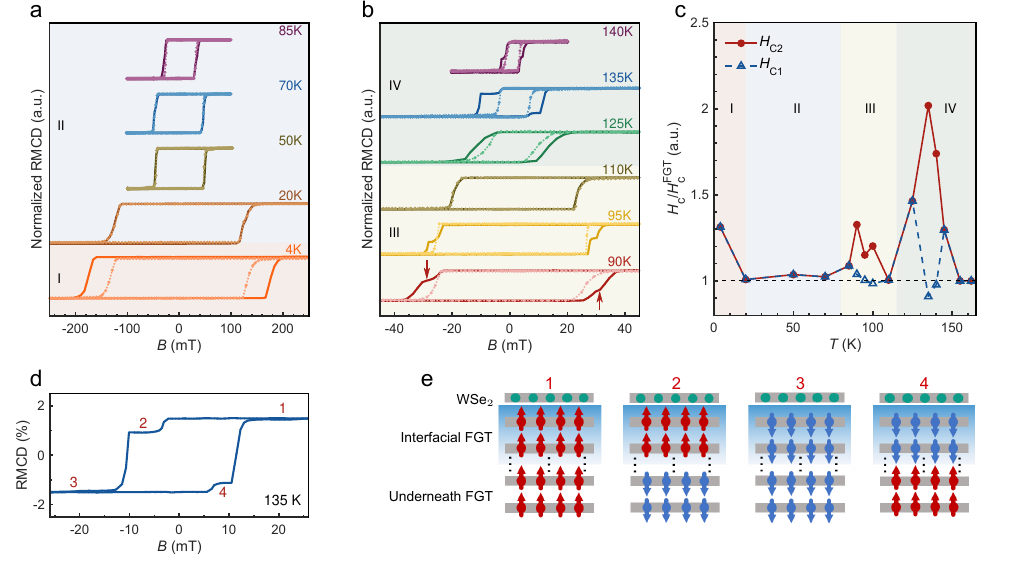}
	\caption{Temperature dependence of the SOC proximity effect in H2. (a,b) Comparison of normalized RMCD signal for bare FGT (dashed curves) and WSe$_2$-covered FGT (solid curves) at several temperatures. (c) The ratio of $H_\mathrm{c}$ in WSe$_2$-covered FGT to that in bare FGT as a function of temperature. Separated parts of two curves correspond to the emergence of the magnetic intermediate states. (d) RMCD signal for WSe$_2$-covered FGT at $135\unit{K}$. (e) Schematics of moment configurations for the corresponding magnetic states in (d). The layer with solid green circles represents the monolayer WSe$_2$, and the solid circles with arrows represent the magnetization orientations of different FGT layers. The gradient blue background depicts the influence induced by SOC proximity.}
	\label{fig5}
\end{figure*}

To explore more details of this interfacial SOC proximity effect, we fabricated another heterostructure H2 for RMCD measurements, consisting of a monolayer WSe$_2$ and a $7.8\unit{nm}$-thick FGT layer. A thin h-BN was used to covered the heterostructure as a comparison.
Figures \ref{fig5}a,b show the normalized RMCD loops at several fixed temperatures (solid curves for WSe$_2$-covered FGT and dashed curves for bare FGT). As the temperature increases, H2 generates four distinct regions marked by Roman numerals. In region I, the enhancement of $H_\mathrm{c}$ is present but diminishes above $20\unit{K}$ in region II. No observable difference of $H_\mathrm{c}$ persists until $85\unit{K}$. In region III, the enhanced $H_\mathrm{c}$ recovers at $90\unit{K}$, accompanied by the emergence of a two-step magnetization reversal (as marked by red arrows in Fig. \ref{fig5}b). Unlike the rectangle loop in the bare FGT, the WSe$_2$-covered FGT exhibits an evident two-step hysteresis loop at about $95\unit{K}$, with two distinct coercive fields $H_\mathrm{c1}$ and $H_\mathrm{c2}$ ($H_\mathrm{c1}<H_\mathrm{c2}$) during the magnetization process.
The two-step loop is unstable at elevated temperatures, vanishing at $110\unit{K}$ but reappearing in region IV. We depict the modulation as a function of temperature in Fig. \ref{fig5}c, defining the modulation strength as the ratio of $H_\mathrm{c}$ in the WSe$_2$-covered FGT to that in the bare FGT. The two-step loop is only observable with separate coercive fields at higher temperatures.
In addition, the enhancement of $H_\mathrm{c}$ in H2 with hBN encapsulation confirms that the altered magnetic order results from the interfacial proximity effect rather than the degeneration (see Figure S3).

The two-step magnetization reversal in the ferromagnetic layer can be achieved by spin exchange coupling with an adjacent antiferromagnetic layer\cite{two-step1,two-step2}. However, this is not the scenario in this work as WSe$_2$ is a nonmagnetic semiconductor. 
Figure \ref{fig5}d shows the two-step hysteresis loop manifesting four stable magnetic states, labeled by red numbers from 1 to 4. The additional intermediate states (state 2 and state 4) could be explained by the short-range nature of the SOC proximity effect. Assuming that the FGT layers fall into two parts due to the proximity of WSe$_2$ (see the schematics in Fig. \ref{fig5}e). For the FGT layers at the interface, the SOC proximity effect is dominant, leading to enhanced magnetic anisotropy and a larger coercive field. While the underneath FGT layers are minimally affected by the interface, resulting in a smaller magnetic field required for flipping. 
Thus the arrangement of magnetic moments can be illustrated by schematics in Fig. \ref{fig5}e. States 1 and 3 exhibit fully spin-polarized magnetization with all-spin-up and all-spin-down configurations, respectively. While the magnetizations in the layers adjacent to and far away from the interface point in opposite directions, resulting in intermediate states 2 and 4. A similar two-step hysteresis loop induced by SOC proximity was recently observed in the MoS$_2$/FGT heterostructure\cite{2022MoS2-FGT}.

At low temperatures, characterized by reduced thermal fluctuation, the interlayer exchange coupling is sufficiently strong to maintain a uniform $H_\mathrm{c}$ in FGT layers. However, the interlayer exchange coupling diminishes with strong thermal fluctuation at elevated temperatures, resulting in non-uniform coercive fields for the interfacial and underneath FGT layers, i.e., the two-step magnetization reversal process. 
At much lower temperatures ($<20\unit{K}$), the proximity effect exhibits a non-monotonic modification depending on the thickness of the WSe$_2$ and the FGT layers. The enhancement of $H_\mathrm{c}$ emerges in monolayer-WSe$_2$/FGT, but disappears in others. While the reduced $H_\mathrm{c}$ is observed with the thinnest FGT ($6.2\unit{nm}$). This may suggest a complex competition between different mechanisms including interlayer exchange coupling, SOC proximity, interfacial strain, thermal fluctuation and so on, which requires further investigation in the future.

In conclusion, we systematically investigated the proximity effect on 2D magnetism in WSe$_2$/FGT vdW heterostructures. Both $H_\mathrm{c}$ and $T_\mathrm{c}$ can be noticeably enhanced in the WSe$_2$-covered region, which is more pronounced as the FGT thickness decreases. This enhancement is attributed to the SOC proximity effect induced from the adjacent WSe$_2$. At low temperatures, we observed a reduction of the coercive field in the heterostructure with a $6.2\unit{nm}$ thick FGT flake, potentially due to the strain effect from the interface.
In addition, the presence of unconventional magnetic stratification demonstrates the short-range nature of the SOC proximity effect. Our findings offer a practical method for modifying 2D magnetic order by combining a 2D nonmagnetic semiconductor with strong SOC, thereby promising for advancing the field of spintronics based on vdW magnetic heterostructures.

\ack Financial support from the National Key R\&D Program of China (Nos. 2022YFA120470, 2021YFA1400400), the National Natural Science Foundation of China (Nos. 12004173, 11974169), and the Fundamental Research Funds for the Central Universities (Nos. 020414380087, 020414913201) are gratefully acknowledged.

\bibliography{IOP_2D}

\providecommand{\newblock}{}
\begin{thebibliography}{10}
\expandafter\ifx\csname url\endcsname\relax
  \def\url#1{{\tt #1}}\fi
\expandafter\ifx\csname urlprefix\endcsname\relax\def\urlprefix{URL }\fi
\providecommand{\eprint}[2][]{\url{#2}}

\bibitem{1990Science}
Prinz G~A 1990 {\em Science\/} {\bf 250} 1092--1097

\bibitem{2004Rev-of-MP}
$\mathrm{\check{Z}}$uti$\mathrm{\acute{c}}$ I, Fabian J and Das~Sarma S 2004
  {\em Rev. Mod. Phys.\/} {\bf 76} 323--410

\bibitem{Datta-Das}
Datta S and Das B 1990 {\em Appl. Phys. Lett.\/} {\bf 56} 665--667

\bibitem{2023AM10}
Moodera J~S, Kinder L~R, Wong T~M and Meservey R 1995 {\em Phys. Rev. Lett.\/}
  {\bf 74} 3273--3276

\bibitem{2023AM11}
Miyazaki T and Tezuka N 1995 {\em J. Magn. Magn. Mater.\/} {\bf 139} L231--L234

\bibitem{2020NPJreview}
Ahn E~C 2020 {\em npj 2D Mater. Appl.\/} {\bf 4} 17

\bibitem{2021NNreview}
Sierra J~F, Fabian J, Kawakami R~K, Roche S and Valenzuela S~O 2021 {\em Nat.
  Nanotech.\/} {\bf 16} 856--868

\bibitem{2021APRreview}
Zhang L, Zhou J, Li H, Shen L and Feng Y~P 2021 {\em Appl. Phys. Rev.\/} {\bf
  8} 021308

\bibitem{2023AMreview}
Ghising P, Biswas C and Lee Y~H 2023 {\em Adv. Mater.\/} {\bf 35} 2209137

\bibitem{Kreview2016}
Novoselov K~S, Mishchenko A, Carvalho A and Castro~Neto A~H 2016 {\em
  Science\/} {\bf 353} aac9439

\bibitem{Duanreview2016}
Liu Y, Weiss N~O, Duan X, Cheng H~C, Huang Y and Duan X 2016 {\em Nat. Rev.
  Mater.\/} {\bf 1} 16042

\bibitem{2dapp3}
Liu Y, Huang Y and Duan X 2019 {\em Nature\/} {\bf 567} 323--333

\bibitem{2017WSe-CrI}
Zhong D, Seyler K~L, Linpeng X, Cheng R, Sivadas N, Huang B, Schmidgall E,
  Taniguchi T, Watanabe K, McGuire M~A, Yao W, Xiao D, Fu K~M~C and Xu X 2017
  {\em Sci. Adv.\/} {\bf 3} e1603113

\bibitem{2020WSe-CrI}
Zhong D, Seyler K~L, Linpeng X, Wilson N~P, Taniguchi T, Watanabe K, McGuire
  M~A, Fu K~M~C, Xiao D, Yao W and Xu X 2020 {\em Nat. Nanotech.\/} {\bf 15}
  187--191

\bibitem{Gating-theory}
Zollner K, Faria~Junior P~E and Fabian J 2019 {\em Phys. Rev. B\/} {\bf 100}
  085128

\bibitem{pr-material}
Li L, Jiang S, Wang Z, Watanabe K, Taniguchi T, Shan J and Mak K~F 2020 {\em
  Phys. Rev. Mater.\/} {\bf 4} 104005

\bibitem{2020MoSe-CrBr}
Lyons T~P, Gillard D, Molina-Sánchez A, Misra A, Withers F, Keatley P~S,
  Kozikov A, Taniguchi T, Watanabe K, Novoselov K~S, Fernández-Rossier J and
  Tartakovskii A~I 2020 {\em Nat. Commun.\/} {\bf 11} 6021

\bibitem{2021ACSnano}
Kim S~J, Choi D, Kim K~W, Lee K~Y, Kim D~H, Hong S, Suh J, Lee C, Kim S~K, Park
  T~E and Koo H~C 2021 {\em ACS Nano\/} {\bf 15} 16395--16403

\bibitem{2022MoS2-FGT}
Tu Z, Zhou T, Ersevim T, Arachchige H~S, Hanbicki A~T, Friedman A~L, Mandrus D,
  Ouyang M, Žutić I and Gong C 2022 {\em Appl. Phys. Lett.\/} {\bf 120}
  043102

\bibitem{WS2-FePS3}
Gong C, Zhang P, Norden T, Li Q, Guo Z, Chaturvedi A, Najafi A, Lan S, Liu X,
  Wang Y, Gong S~J, Zeng H, Zhang H, Petrou A and Zhang X 2023 {\em Nat.
  Commun.\/} {\bf 14} 3839

\bibitem{2023ACSnano}
Choi E~M, Kim T, Cho B~W and Lee Y~H 2023 {\em ACS Nano\/} {\bf 17}
  15656--15665

\bibitem{2dmag_CrI3}
Huang B, Clark G, Navarro-Moratalla E, Klein D~R, Cheng R, Seyler K~L, Zhong D,
  Schmidgall E, McGuire M~A, Cobden D~H, Yao W, Xiao D, Jarillo-Herrero P and
  Xu X 2017 {\em Nature\/} {\bf 546} 270--273

\bibitem{fgt2018}
Fei Z, Huang B, Malinowski P, Wang W, Song T, Sanchez J, Yao W, Xiao D, Zhu X,
  May A~F, Wu W, Cobden D~H, Chu J~H and Xu X 2018 {\em Nat. Mater.\/} {\bf 17}
  778--782

\bibitem{SongC-PRB17}
Wang H, Liu Y, Wu P, Hou W, Jiang Y, Li X, Pandey C, Chen D, Yang Q, Wang H,
  Wei D, Lei N, Kang W, Wen L, Nie T, Zhao W and Wang K~L 2020 {\em ACS Nano\/}
  {\bf 14} 10045--10053

\bibitem{SongC-PRB20}
Katmis F, Lauter V, Nogueira F~S, Assaf B~A, Jamer M~E, Wei P, Satpati B,
  Freeland J~W, Eremin I, Heiman D, Jarillo-Herrero P and Moodera J~S 2016 {\em
  Nature\/} {\bf 533} 513--516

\bibitem{W-CGT}
Zhu W, Song C, Han L, Bai H, Wang Q, Yin S, Huang L, Chen T and Pan F 2022 {\em
  Adv. Funct. Mater.\/} {\bf 32} 2108953

\bibitem{CGT-SOC}
Dong X~J, You J~Y, Zhang Z, Gu B and Su G 2020 {\em Phys. Rev. B\/} {\bf 102}
  144443

\bibitem{WSe-SOC-theory}
Le D, Barinov A, Preciado E, Isarraraz M, Tanabe I, Komesu T, Troha C, Bartels
  L, Rahman T~S and Dowben P~A 2015 {\em J. Phys.: Condens. Matter\/} {\bf 27}
  182201

\bibitem{PDMS_method}
Castellanos-Gomez A, Buscema M, Molenaar R, Singh V, Janssen L, van~der Zant
  H~S~J and Steele G~A 2014 {\em 2D Mater.\/} {\bf 1} 011002

\bibitem{WSe-PL}
Zhao W, Ghorannevis Z, Chu L, Toh M, Kloc C, Tan P~H and Eda G 2013 {\em ACS
  Nano\/} {\bf 7} 791--797

\bibitem{FGT-PRB}
Zhuang H~L, Kent P~R~C and Hennig R~G 2016 {\em Phys. Rev. B\/} {\bf 93} 134407

\bibitem{FGTstrain}
Wang Y, Wang C, Liang S~J, Ma Z, Xu K, Liu X, Zhang L, Admasu A~S, Cheong S~W,
  Wang L, Chen M, Liu Z, Cheng B, Ji W and Miao F 2020 {\em Adv. Mater.\/} {\bf
  32} 2004533

\bibitem{two-step1}
Chen P, Huang Z, Li C, Zhang B, Bao N, Yang P, Yu X, Zeng S, Tang C, Wu X, Chen
  J, Ding J, Pennycook S~J, Ariando A, Venkatesan T~V and Chow G~M 2018 {\em
  Adv. Funct. Mater.\/} {\bf 28} 1801766

\bibitem{two-step2}
Wu X, Lan D, Hwang I, Sun C, Zhou H, Yu X, Yang P, Yu X, Liu C, Chen P, Ding J,
  Chen J and Chow G~M 2023 {\em J. Alloy. Compd.\/} {\bf 932} 167582

\end{thebibliography}
\bibliographystyle{iopart-num}

\end{document}